# High Accuracy Phishing Detection Based on Convolutional Neural Networks

Suleiman Y. Yerima[1]    and    Mohammed K. Alzaylaee[2]

[1]Cyber Technology Institute
Faculty of Computing, Engineering and Media,
De Montfort University, Leicester, United Kingdom
syerima@dmu.ac.uk

[2]Al-Qunfudah College of Computing,
Umm Al-Qura University, Saudi Arabia
mkzaylaee@uqu.edu.sa

*Abstract*— The persistent growth in phishing and the rising volume of phishing websites has led to individuals and organizations worldwide becoming increasingly exposed to various cyber-attacks. Consequently, more effective phishing detection is required for improved cyber defence. Hence, in this paper we present a deep learning-based approach to enable high accuracy detection of phishing sites. The proposed approach utilizes convolutional neural networks (CNN) for high accuracy classification to distinguish genuine sites from phishing sites. We evaluate the models using a dataset obtained from 6,157 genuine and 4,898 phishing websites. Based on the results of extensive experiments, our CNN based models proved to be highly effective in detecting unknown phishing sites. Furthermore, the CNN based approach performed better than traditional machine learning classifiers evaluated on the same dataset, reaching 98.2% phishing detection rate with an F1-score of 0.976. The method presented in this paper compares favourably to the state-of-the art in deep learning based phishing website detection.

*Keywords—Phishing; Convolutional Neural Networks; Machine learning; Deep learning; Phishing website detection; Social Engineering*

## I. INTRODUCTION

Phishing is a social engineering based attack which enables cybercriminals to steal credentials, distribute ransomware, and carry out financial fraud and theft. It also enables nation-state actors to gain strategic access to target environments. Through well-designed counterfeit websites, phishing is used to obtain private sensitive information, e.g. account number and password from unsuspecting users. The 2019 Phishlabs trends and intelligence report [1] states that phishing grew by 40.9% in 2018 with 83.9% of the observed attacks targeting credentials for financial, email, cloud, payment and SaaS services. According to the report, the volume of phishing websites (i.e. phishing content located on a unique fully qualified domain name or host) rose steadily during the first quarter of 2018 and remained high throughout the second and third quarters. Furthermore, the total number of phishing sites observed monthly significantly surpassed previous years.

The need for effective countermeasures has made phishing detection a popular area of research in recent years. Consequently, three main categories of approaches for phishing detection have emerged: (a) Approaches based on blacklists and whitelists [2], [3] (b) Approaches based on web page visual similarity [4] (c) Approaches based on URL and website content features [5]. The blacklist approach is ineffective in detecting new phishing websites that the system has not yet been updated with. The visual similarity-based method extracts visual features from phishing websites, and then uses these features to identify phishing webpages. Hence, any distortion of web page content affects the visual content retrieval leading to misclassification. Most current phishing detection approaches exploit the URL and web content features to distinguish between phishing and genuine websites e.g. [5], [6]. Machine learning techniques have also been integrated with URL and web content features to improve detection performance and enable zero-day phishing defence e.g. [7-9].

Given the persistent growth in phishing attacks and the steady rise in phishing sites, the need for more effective means of detecting suspect sites and thwarting zero-day phishing attacks has never been greater. Hence, in this paper we present a deep learning based approach that utilizes Convolutional Neural Networks (CNN) for high accuracy phishing website detection. Our approach exploits URL and web contents features to build machine learning based phishing detection models that are capable of detecting new, previously unseen phishing websites.

We present the design of our CNN-based model for phishing website detection and evaluate the model on a dataset obtained from 4,898 phishing websites and 6,157 genuine websites [10], [14]. Furthermore, we compare the performance of our CNN model to other popular machine learning classifiers including Naïve Bayes, Bayes Net, Decision Tree, SVM, Random Forest, Random Tree and Simple Logistic on the same dataset. The comparative analysis shows that the CNN-based model ultimately achieves the best phishing detection performance of 98.2% with an F1-score 0f 0.976.

The rest of the paper is organized as follows: Related work is in Section II; Background on CNN is featured in Section III; Section IV presents methodology and the experiments performed; Results of experiments are given in Section V and finally Section VI presents the conclusions of the study and future work.

## II. RELATED WORK

Phishing detection based on machine learning is a growing field of study with increasing interest in application of deep learning techniques. Yuan et al [9], proposed a method based on features from URLs and web page links to detect phishing website and their targets. They utilized a Deep Forest model that results in a true positive rate of 98.3% and a false alarm rate of 2.6%. In particular, they designed an effective strategy based on search operator via search engines to find the phishing targets, which achieves an accuracy of 93.98%.

Wang, et al. [8] presented PDRCNN, a phishing website detection approach that utilizes only the URL of the website to build detection models. Their system combines RNNs and CNN to extract features from the URL strings. In their experiments, detection accuracy of 97% and AUC of 99% were achieved. Bahnsen et al. [15] presented an LSTM model to detect phishing URLs. Their approach first encodes the URL strings using one-hot encoding and then inputs each encoded character vector into the LSTM neurons for training and testing. Their method achieved an accuracy of 0.935 on the Common Crawl and PhishTank datasets. Hung et al. [16], presented the URLNet method for malicious website URL detection. They extracted both character level and word-level features based on URL strings and utilized Convolutional Neural Network for training and testing. The authors of [17] propose and evaluate a system that extracts features from URL using Natural Language Processing (NLP) techniques. Their system was implemented by examining URLs used in phishing attacks and extracting the features from them. The authors tested their system on several machine learning algorithms and found Random Forest to have the best performance with a success rate of 89.9%. The drawback of the URL only approach is that correct classification may not be obtained if the URL itself lacks the relevant semantics, or if there is a problem with the validity of the URL [8]. The CNN based approach presented in this paper utilizes not only URL features but features from other properties of the websites, which increases robustness.

In [18], the authors propose a hybrid intelligent phishing website prediction system using deep neural networks (DNN) with evolutionary algorithm-based feature selection and weighting methods for enhanced prediction. Genetic Algorithm (GA) is used to heuristically identify the most influential features and optimal weights of the website features. Unlike the study in [18], our approach requires no feature selection stage as this is implicitly performed within the CNN-based model due to its design. Other deep learning-based phishing detection works include [19],[23], with some studies extracting features from emails [20-22] rather than URLs and webpage characteristics.

## III. BACKGROUND

### A. Convolutional Neural Networks (CNN)

CNN belongs to the family of Artificial Neural Networks which are computational models inspired by the characteristics of biological neural networks. A CNN is a deep learning technique that works well for identifying simple patterns in the data which will then be used to form more complex patterns in subsequent layers. Two types of layers are typically used for building CNNs; convolutional layers and pooling layers. The role of the convolutional layer is to detect local conjunctions of features from the previous layer, while the role of the pooling layer is to merge semantically similar features into one [11].

Generally, the convolutional layer extracts the optimal features, the pooling layer reduces the dimensions of the convolutional layer features, and fully connected layer(s) are then used for classification. The performance of the CNN is generally influenced by the number of layers and the number of filters (kernels). More and more abstract features are extracted in the deeper layers of the CNN, hence, the number of layers required depends on the complexity and non-linearity of the data being analysed. Furthermore, the number of filters in each stage determines the number of features extracted. Computational complexity increases with more layers and higher numbers of filters. Also, with more complex architectures there is the possibility of training an overfitted model which results in poor prediction accuracy on the testing set(s). To reduce overfitting, techniques such as 'dropout' [12] and 'batch regularization' are implemented during training of our models.

### B. One Dimensional Convolutional Neural Networks

Although CNN is more commonly applied in a multi-dimensional fashion and has thus found success in image and video analysis-based problems, they can also be applied to one-dimensional data. Datasets that possess a one-dimensional structure can be processed using a one-dimensional convolutional neural network (1D CNN). The key difference between a 1D and a 2D or 3D CNN is the dimensionality of the input data and how the filter (feature detector) slides across the data. For 1D CNN, the filters only slide across the input data in one direction. A 1D CNN is quite effective when you expect to derive interesting features from shorter (fixed-length) segments of the overall dataset, and where the location of the feature within the segment is not of high relevance.

The use of 1D CNN can be commonly found in NLP applications. Similarly, 1D CNN is applicable to datasets containing vectorised data being used to characterize the items to be predicted (e.g. a website). The 1D CNN could be used to extract potentially more discriminative feature representations that describe any existing patterns or relationships within segments of the vectors characterizing each entity in the dataset. These new features are then fed into a classifier (e.g. a fully connected neural network layer) which will in turn use the derived features in making a final classification decision. Hence, in this scenario, the convolutional layers can be considered as a feature extractor that eliminates the need for feature ranking and selection. The CNN model developed in this paper is applied to vectorised data characterizing the websites in order to derive a trained model that can detect new phishing websites with very high accuracy.

*C. Key elements of our proposed CNN architecture*

Our proposed CNN architecture is a 1D CNN consisting of two convolutional layers and two max pooling layers. These are followed by a Fully Connected layer of *N* units, which is in turn connected to a final classification layer containing one neuron with a *sigmoid* activation function.

The sigmoid activation function is given by: $S = \frac{1}{1+e^{-x}}$

The final classification later generates an outcome corresponding to the two classes i.e. 'Phishing' or 'Legitimate'. The convolutional layers utilize the *ReLU* (Rectified Linear Units) activation function given by: $f(x) = \max(0, x)$. *ReLU* helps to mitigate vanishing and exploding gradient issues [13]. It has been found to be more efficient in terms of time and cost for training huge data in comparison to classical non-linear activation functions such as Sigmoid or Tangent functions [13]. A simplified view of our architecture is shown in Figure 1.

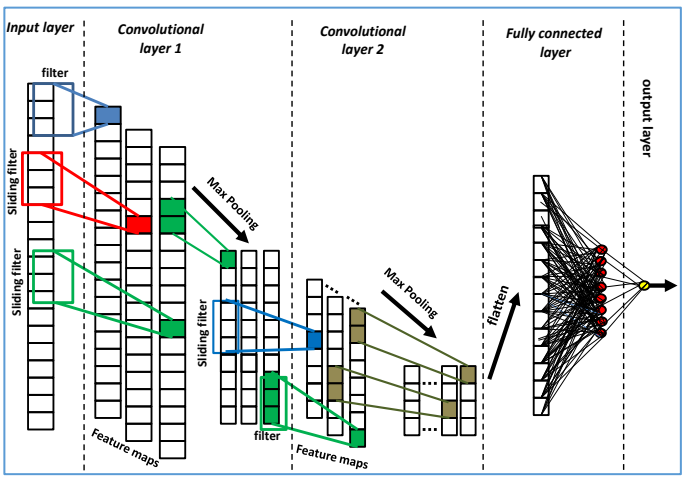

Figure 1: Simplified view of the implemented 1D CNN model for phishing website detection.

## IV. METHODOLOGY AND EXPERIMENTS

In this section we present the experiments undertaken to evaluate the CNN models developed in this paper. Our models were implemented using Python and utilized the Keras library with TensorFlow backend. Other libraries used include Scikit Learn, Seaborn, Pandas, and Numpy. The model was built and evaluated on an Ubuntu Linux 16.04 64-bit Machine with 4GB RAM.

*A. Problem definition*

Let W = $\{w_1, w_2, \ldots w_n\}$ be a set of website samples where each $w_i$ is represented by a vector containing the values of *f* attributes (as shown in Table 1). Let $w_i = \{a_1, a_2, a_3 \ldots a_f, cl\}$ where $cl \in \{Phishing, Legitimate\}$ is the class label assigned to the website. Thus, W can be used to train the model to learn the behaviours of Phishing and Legitimate websites respectively. The goal of a trained model is then to classify a given unlabelled website $w_{unknown} = \{a_1, a_2, a_3 \ldots a_f, ?\}$ by assigning a label *cl*, where $cl \in \{Phishing, Legitimate\}$.

*B. Dataset*

In our experiments we used the benchmarked dataset from [10]. Detailed descriptions of the features/attributes in the dataset can be found in [5], [10] and [14]. Table 1 presents a summary of the attributes. The dataset consists of 11,055 instances obtained from 4,898 phishing websites and 6,157 legitimate websites.

Table 1: Features of the phishing and legitimate websites in dataset.

| Attribute number | Attributes | Possible values |
|---|---|---|
| 1 | having_IP_Address | -1,1 |
| 2 | URL_Length | 1,0,-1 |
| 3 | Shortening_Service | 1,-1 |
| 4 | having_At_Symbol | 1,-1 |
| 5 | double_slash_redirecting | -1,1 |
| 6 | Prefix_Suffix | -1,1 |
| 7 | having_Sub_Domain | -1,0,1 |
| 8 | SSLfinal_State | -1,1,0 |
| 9 | Domain_registration_length | -1,1 |
| 10 | Favicon | 1,-1 |
| 11 | Port | 1,-1 |
| 12 | HTTPS_token | -1,1 |
| 13 | Request_URL | -1,1 |
| 14 | URL_of_Anchor | -1,0,1 |
| 15 | Links_in_tags | 1,-1,0 |
| 16 | SFH (server form handler) | -1,1,0 |
| 17 | Submitting_to_email | -1,1 |
| 18 | Abnormal_URL | -1,1 |
| 19 | Redirect page | 0,1 |
| 20 | onMouseOver ( using to hide link) | 1,-1 |
| 21 | RightClick | 1,-1 |
| 22 | Using pop-up widnow | 1,-1 |
| 23 | Iframe | 1,-1 |
| 24 | age_of_domain | -1,1 |
| 25 | DNSRecord | -1,1 |
| 26 | web_traffic | -1,0,1 |
| 27 | Page_Rank | -1,1 |
| 28 | Google_Index | -1,1 |
| 29 | Links_pointing_to_page | 1,0,-1 |
| 30 | Statistical_report | -1,1 |
| Class | Result | -1,1 |

*C. Experiments to evaluate the proposed CNN based model*

In order to investigate the performance of our proposed model we performed different sets of experiments. The first set of experiments was aimed at evaluating the impact of different number of layers on the model's performance. Table 2 shows the configurations of the CNN models. CNN1 consists of 1 convolutional layer, followed by a max pooling layer. The output of the max pooling layer is flattened and passed on to a fully connected layer with 8 units. This is in turn connected to a sigmoid activated output layer containing one unit. CNN2 has the same configuration but with two sets of convolutional and max pooling layer as shown in Table 2. In this set of experiments the number of filters (kernels) was also varied to examine the impact on the performance of the models.

The second set of experiments involved varying the length of filters (i.e. kernel size) while keeping the number of filters fixed. The results of the experiments are discussed in the next section. In order to measure model performance, we used the following metrics: *Accuracy, precision, recall and F1-score*. The metrics are defined as follows.

Table 2: Summary of model configurations used in the experiments.

| Model design summary | |
|---|---|
| CNN1 | **1D Convolutional layer:** 8,16, 32, 64 filters, size = 10<br>**MaxPooling layer:** Size =2, Stride = 2<br>**Fully Connected layer:** 8 units, activation=ReLU<br>**Output layer: Fully Connected layer**: 1 unit, activation=sigmoid |
| CNN2 | **1D Convolutional layer:** 8, 16, 32, 64 filters, size = 10<br>**MaxPooling layer:** Size =2, Stride = 2<br>**1D Convolutional layer:** 8, 16, 32, 64 filters, size = 5<br>**MaxPooling layer:** Size =2, Stride = 2<br>**Fully Connected layer: 8 units**, activation=ReLU<br>**Output layer: Fully Connected layer**; 1 unit, activation=sigmoid |

- *Accuracy*: Defined as the ratio between correctly predicted outcomes and the sum of all predictions. It is given by: $\frac{TP+TN}{TP+TN+FP+FN}$
- *Precision*: All true positives divided by all positive predictions. i.e. Was the model right when it predicted positive? Given by: $\frac{TP}{TP+FP}$
- *Recall*: True positives divided by all actual positives. I.e. how many positives did the model identify out of all possible positives? Given by: $\frac{TP}{TP+FN}$
- *F1-score*: This is the weighted average of precision and recall, given by: $\frac{2 \times Recall \times Precision}{Recall+Precision}$

Where TP is true positives; FP is false positives; FN is false negatives, while TN is true negatives (w.r.t. the Phishing class). All the results of the experiments are from 10-fold cross validation where the dataset is divided into 10 equal parts with 10% of the dataset held out for testing, while the models are trained from the remaining 90%. This is repeated until all of the 10 parts have been used for testing. The average of all 10 results is then taken to produce the final result. Also, during the training of the CNN models (for each fold), 10% of the training set was used for validation.

## V. RESULTS AND DISSCUSSIONS

### A. Impact of number of layers and numbers of filters.

In this section we examine the results from CNN1 and CNN2 respectively. Table 3 shows the results from running the CNN1 with different numbers of filters. Table 4 contains the results of CNN2 with different numbers of filters. From Table 3, it is evident that the number of filters had an effect on the performance of the CNN1 model. With a larger number of filters (32, 64) a higher accuracy of 96.6% is observed. While an F1-score of 0.97 is observed with the higher number of (32, 64) filters. As mentioned earlier, the number of filters indicates the number of features being extracted, with more filters increasing the complexity of the model and hence, more parameters to train. Note that with the CNN1 model, similar performance is obtainable with 32 filters and 64 filters while requiring to train 2969 vs. 5,913 parameters respectively, as seen from Table 3. The overall accuracy for 8 filters and 16 filters are 95.8% and 96.2% respectively. The length of filters used in each case was fixed at 10.

Table 3: 1-layer CNN results (length of filters used =10)

| Number of Filters | 8 | 16 | 32 | 64 |
|---|---|---|---|---|
| Accuracy | 0.958 | 0.962 | 0.966 | **0.966** |
| Precision | 0.958 | 0.967 | 0.967 | 0.965 |
| Recall | 0.967 | 0.964 | 0.972 | **0.975** |
| F1-score | **0.963** | 0.966 | 0.970 | **0.970** |
| Train Time (s) | 209.76 | 267.22 | 363.86 | 643.15 |
| Test Time (s) | 0.36 | 0.426 | 0.377 | 0.680 |
| Loss | 0.107 | 0.098 | 0.092 | 0.094 |
| Total/ Trainable parameters | 793/ 761 | 1545/ 1497 | 3049/ 2969 | 6,057/ 5,913 |

Table 4: 2- layer CNN results (length of filters used= 10 for first layer and =5 for second layer)

| Number of filters | 8 | 16 | 32 | 64 |
|---|---|---|---|---|
| Accuracy | 0.958 | 0.964 | 0.969 | **0.971** |
| Precision | 0.959 | 0.961 | 0.967 | 0.966 |
| Recall | 0.967 | 0.975 | 0.978 | **0.983** |
| F1-score | 0.963 | 0.968 | 0.972 | **0.974** |
| Train Time (s) | 373.62 | 334.54 | 453.6 | 804.89 |
| Test Time (s) | 0.56 | 0.553 | 0.497 | 0.615 |
| loss | 0.11 | 0.099 | 0.1 | 0.098 |
| Total/ Trainable parameters | 697/ 649 | 1993/ 1913 | 6,361/ 6,505 | 23,209/ 22,937 |

From Table 4, it is evident that two sets of convolutional and max pooling layers (CNN2) results in improvement over CNN1. However, the margin of improvement suggests that adding another CNN layer (to make it a 3-layer CNN model) is unlikely to significantly improve performance, whereas the number of parameters to be trained will increase dramatically. As with CNN1, the number of filters used improves performance in the CNN2 model. The use of 64 filters resulted in the highest accuracy of 97.1% with F1-score of 0.974 compared to 95.8% accuracy and 0.963 F1-score obtained with only 8 filters.

*1) Training epochs, loss and accuracy graphs.*

Figures 2 and 3 shows the typical outputs obtained with the validation and training sets during the training epochs up to 220 epochs. From Fig. 2, the training and validation accuracies matched up to each other quite closely, indicating that the training was not overfitting the model to the training set. Figure 3 shows the typical loss behaviour observed during the experiments. The training and validation losses also followed one another quite closely. In order to increase the possibility of obtaining the 'best' trained model and reduce training time we implemented a 'stopping criterion' which will stop the training once no improvement in performance is observed within 50 epochs.

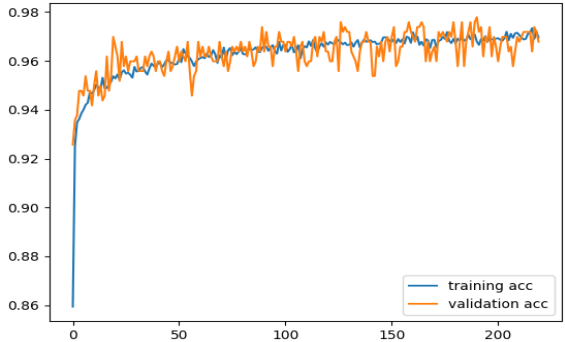

Figure 2: Training and validation accuracies at different epochs up to 220, for the CNN model.

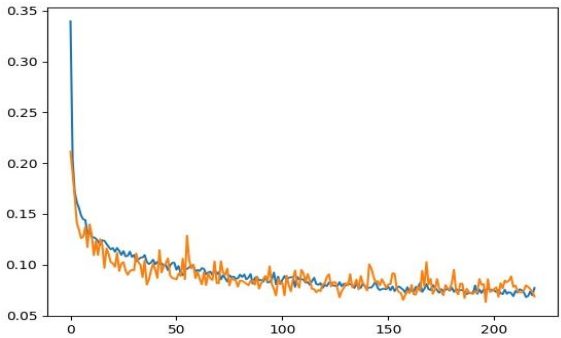

Figure 3: Training and validation losses at different epochs up to 220, for the CNN model

*B. Impact of the length of filters on performance.*

In this section we examine the effect of the length of filters by using the CNN2 model with the number of filters fixed at 64. The length is varied from 4, 6, 8, 10 to 12 respectively. The results indicate that the highest accuracy of 97.2% is attained with an F1-score of 0.975 when the filter length for the first convolutional layer is set at 12. Recall that in the design of our model, the length of filters for the second convolutional layer is set to half that of the first layer.

From Table 5, it can be seen that a filter length of 12 (with 6 in the second convolutional layer) achieved an overall accuracy of 97.2% and F1-score of 0.975; compared to a filter length of 4 which achieved overall accuracy of 96.7% and F1-score of 0.970.

Table 5: Length of filters (CNN2); number of filters =64.

| Length of filters | 4 | 6 | 8 | 10 | 12 |
|---|---|---|---|---|---|
| Accuracy | 0.967 | 0.969 | 0.970 | 0.971 | **0.972** |
| Precision | 0.964 | 0.967 | 0.971 | 0.966 | 0.969 |
| Recall | 0.976 | 0.978 | 0.975 | **0.983** | 0.981 |
| F1-score | 0.970 | 0.972 | 0.973 | 0.974 | **0.975** |
| Train Time (s) | 827.40 | 936.92 | 822.05 | 804.89 | 640.01 |
| Test Time (s) | 0.618 | 0.562 | 0.522 | 0.615 | 0.472 |
| loss | 0.086 | 0.091 | 0.095 | 0.098 | 0.096 |
| Total/ Trainable parameters | 12,073/ 11,801 | 15,785/ 15,513 | 19,497/ 19,225 | 23,209/ 22,937 | 27,985/ 27,697 |

*1) Obtaining optimal performance*

Achieving the optimal performance point for a CNN model is non-trivial due to several parameters that require tuning. We further experimented with different number of units in the fully connected layer, leaving the number of filters at 64 and the length at 12 for first layer and 6 at the second layer. The best result obtained (with the CNN2 model) was the following: **Accuracy: 0.973; Precision: 0.970; Recall: 0.982; F1-score: 0.976.** This was obtained by increasing the number of units in the fully connected layer from 8 units to 32 units. In this configuration the number of trainable parameters only increased to 29,793.

*C. CNN performance vs. other machine learning classifiers: 10 fold cross validation results.*

In Table 6, the performance of CNN architecture developed in this paper is compared to other machine learning classifiers: Naïve Bayes, SVM, Bayes Net, J48, Random Tree, and Random Forest. Figure 4 shows the F1-scores of the classifiers, where CNN has the highest F1-score, followed by Random Forest and Random Tree. Figure 5 depicts the overall accuracy where CNN outperforms six of the classifiers, with Random Forest achieving the same accuracy. Table 6 shows that the recall of CNN is 0.982 which indicates that it has the best phishing website detection rate compared to the other 7 classifiers.

Table 6: Comparison with other ML classifiers.

| | ACC | Prec. | Rec. | F1 |
|---|---|---|---|---|
| **Naïve Bayes** | 0.907 | 0.904 | 0.884 | 0.894 |
| **SVM** | 0.927 | 0.931 | 0.903 | 0.916 |
| **RF** | 0.973 | **0.977** | 0.961 | 0.969 |
| **SL** | 0.928 | 0.932 | 0.904 | 0.918 |
| **J48** | 0.960 | 0.966 | 0.942 | 0.954 |
| **Random Tree** | 0.963 | 0.964 | 0.952 | 0.958 |
| **Bayes Net** | 0.928 | 0.934 | 0.901 | 0.917 |
| **CNN** | **0.973** | 0.970 | **0.982** | **0.976** |

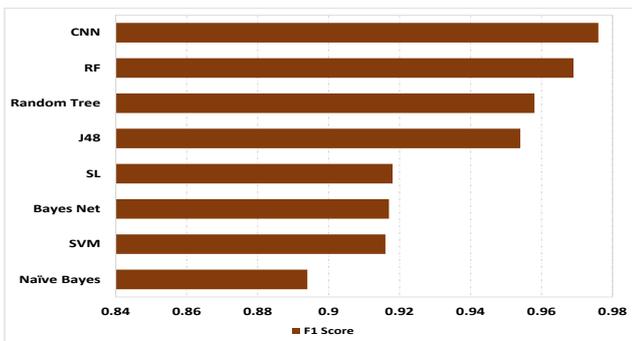

Figure 4: F1-score

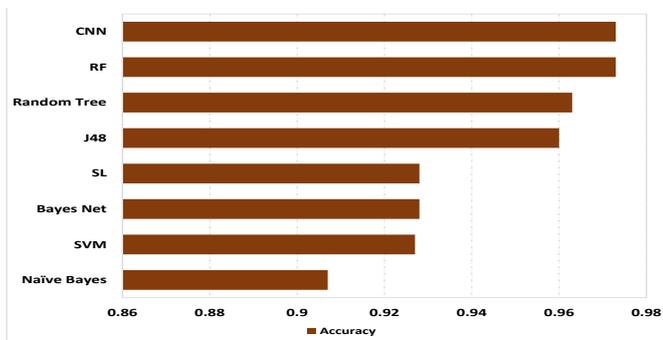

Figure 5: Overall accuracy.

## VI. CONCLUSIONS AND FUTURE WORK

In this paper we proposed a deep learning model based on 1D CNN for the detection of phishing websites. We evaluated the model through extensive experiments on a benchmarked dataset containing 4,898 instances and 6,157 instances from phishing websites and legitimate websites respectively. The model outperforms several popular machine learning classifiers evaluated on the same dataset. The results indicate that our proposed CNN based model can be used to detect new, previously unseen phishing websites more accurately than the other models. For future work, we will aim to improve the model training process by automating the search and selection of the key influencing parametrs (i.e. number of filters, filter lenghts, and number of fully connected units) that jointly results in the optimal performing CNN model.